\documentclass[12pt,preprint]{aastex}

\usepackage{epsfig}
                                                                                
\def\vkm{km s$^{-1}$}

\def\degree{$^\circ$}

\def\arcs#1{$#1''$}
\def\arcsa#1#2{$#1^{\prime\prime}_{^\textrm{.}}#2$}
\def\arcsaq#1#2{#1^{\prime\prime}_{^\textrm{.}}#2}
\def\smassrate{$M_\odot$ yr$^{-1}$}
\def\solarmass{$M_\odot$}

\def\solarlum{$L_\odot$}

\def\Jyb{Jy beam$^{-1}$}
\def\mJyb{mJy beam$^{-1}$}

\def\Tbol{T_\textrm{\scriptsize bol}}

\def\cmc{cm$^{-3}$}

\def\micron{$\mu$m}
                                                                                
\def\ra#1#2#3#4{#1^\mathrm{h} #2^\mathrm{m} #3^\mathrm{s}_{^\textrm{.}} #4}
\def\dec#1#2#3#4{#1\degr #2\arcmin #3^{\prime\prime}_{^\textrm{.}}#4}

\def\Rin{R_\textrm{\scriptsize in}}
\def\Rout{R_\textrm{\scriptsize out}}

\def\H2{H$_2$}
\def\N2HP{N$_2$H$^+$}

\def\cCO{C$^{18}$O}

\def\aCO{$^{12}$CO}
\def\bCO{$^{13}$CO}
\def\NH3{NH$_3$}

\def\SOta{$N_J=5_6-4_5$}

\def\putfig#1#2#3{\epsfig{scale=#1,angle=#2,figure=#3}}
\def\putfig#1#2#3{\includegraphics[angle=#1,scale=#2]{#3}}
\def\putfig#1#2#3{}
\def\leftblank#1{}

\begin{document}

\title{A Rotating Disk in the HH 111 Protostellar System}
\author{Chin-Fei Lee\altaffilmark{1}
}
\altaffiltext{1}{Academia Sinica Institute of Astronomy and Astrophysics,
P.O. Box 23-141, Taipei 106, Taiwan; cflee@asiaa.sinica.edu.tw}

\begin{abstract}
The HH 111 protostellar system is a young Class I system with two sources,
VLA 1 and VLA 2, at a distance of 400 pc. Previously, a flattened envelope has
been seen in \cCO{} to be in transition to a rotationally supported disk
near the VLA 1 source. The follow-up study here is to confirm the
rotationally supported disk at 2$-$3 times higher angular resolutions, at
$\sim$ \arcsa{0}{3} (or 120 AU) in 1.33 mm continuum, and
$\sim$ \arcsa{0}{6} (or 240 AU) in
\bCO{} ($J=2-1$) and \aCO{} ($J=2-1$) emission obtained with the
Submillimeter Array. The 1.33 mm continuum emission shows a resolved dusty
disk associated with the VLA 1 source perpendicular to the jet axis, with a
Gaussian deconvolved size of $\sim$ 240 AU. The \bCO{} and \aCO{} emissions
toward the dusty disk show a Keplerian rotation, indicating that the dusty
disk is rotationally supported. The density and temperature distributions in
the disk derived from a simple disk model are found to be similar to those
found in bright T-Tauri disks, suggesting that the disk can evolve into a
T-Tauri disk in the late stage of star formation. In addition, a hint of a
low-velocity molecular outflow is also seen in \bCO{} and \aCO{} coming out
from the disk.
\end{abstract}


\section{Introduction}

Stars are formed inside molecular cloud cores by means of gravitational
collapse. The details of the process, however, are complicated by the
presence of magnetic fields and angular momentum.
In particular, for the stars to form, collapsing material has to lose 
most of the magnetic fields and angular momentum.
As a result, in addition to
infall (or collapse), rotation and outflow are also seen toward star-forming
regions. In theory, a rotationally supported disk is expected to form
inside a collapsing core around a protostar, from which part of the material
is accreted by the protostar and part is ejected away. Observationally,
however, when and how a rotationally supported disk is actually formed are
still unclear, because of the lack of detailed kinematic studies inside the
collapsing core.

%The source is more exposed due to ejection? 

This paper is a follow-up study of the HH 111 protostellar system, which is
a well-studied Class I system deeply embedded in a compact molecular cloud
core in the L1617 cloud of Orion at a distance of 400 pc. This system has a
bolometric temperature $\Tbol \sim 78$ K and thus has just transitioned to
the Class I phase from the Class 0 phase \citep{Froebrich2005}.  At the
center of this system, there are two sources, VLA 1 and VLA 2, with a
projected separation of $\sim$ \arcs{3} (1200 AU) \citep{Reipurth1999}.
Previously, a flattened envelope has been seen in \cCO{} around the VLA 1
source (Lee et al 2009, hereafter Paper I) and it seems to be in transition
to a rotationally supported disk near the VLA 1 source (Lee et al 2010,
hereafter Paper II). The follow-up study here is to confirm the rotationally
supported disk at 2$-$3 times higher angular resolutions, at
$\sim$ \arcsa{0}{3} (or 120 AU) in 1.33 mm continuum, and
$\sim$ \arcsa{0}{6} (or 240 AU) in
\bCO{} ($J=2-1$) and \aCO{} ($J=2-1$) emission obtained with the
Submillimeter Array (SMA)\footnote{The Submillimeter Array is a joint
project between the Smithsonian Astrophysical Observatory and the Academia
Sinica Institute of Astronomy and Astrophysics, and is funded by the
Smithsonian Institution and the Academia Sinica.}
\citep{Ho2004}. This study also shows that the physical properties of the
disk here in the early phase of star formation can be similar to those of
the bright T-Tauri disks in the late phase of star formation 
\citep{Andrews2009}.

\section{Observations}

Observations toward the HH 111 system were carried out with the SMA, first
on 2005 December 5 in the compact-north configuration
(\citet{Lee2009HH111}), then on 2010 January 24 in the extended
configuration and on 2010 March 26 in the sub-compact configuration
\citep{Lee2010HH111}, and then on 2011 Jan 01 in the very extended
configuration. The baselines have projected lengths ranging from $\sim$ 10
to 480 m onto the sky. The 230 GHz band receivers were used to observe the \aCO{}
($J=2-1$), \bCO{} ($J=2-1$),
\cCO{} ($J=2-1$), and SO (\SOta) lines simultaneously with the 1.33 mm continuum.
The \aCO{} and \bCO{} lines are bright and thus are presented here to show
the kinematics of the inner region and search for the rotationally supported
disk around the VLA 1 source at high resolution. The velocity resolution in
these lines is $\sim$ 0.28 \vkm{} per channel, similar to that of \cCO{}
presented in \citet{Lee2010HH111}. The visibility data were calibrated with
the MIR package. The flux uncertainty was estimated to be $\sim$ 15\%. The
calibrated visibility data in different configurations were combined and
then imaged with the MIRIAD package, as described in
\citet{Lee2010HH111}.
With various different weightings, the synthesized beams can have
a size from $\sim$ \arcsa{0}{6} down to \arcsa{0}{3}.
The rms noise levels are $\sim$ 0.036 \Jyb{} in the 
\aCO{} channel maps with a beam of \arcsa{0}{58}$\times$\arcsa{0}{50},
0.032 \Jyb{} in the \bCO{} channel maps with a beam of 
\arcsa{0}{66}$\times$\arcsa{0}{56},
and 0.9 \mJyb{} in the continuum map with a beam of
\arcsa{0}{36}$\times$\arcsa{0}{31}.
The velocities in the channel maps are LSR.

\section{Results}

% updated with the refined positions of the phase calibrators 

As in \citet{Lee2010HH111}, the results are presented in comparison to a
mosaic image based on the Hubble Space Telescope (HST) NICMOS image ([FeII]
1.64 \micron{} + \H2{} at 2.12 \micron{} + continuum) obtained by
\citet{Reipurth1999}, which shows the two sources VLA 1 and VLA 2 in the
infrared, and the reflection nebulae that trace the illuminated outflow cavity
walls. The two sources are located at
$\alpha_{(2000)}=\ra{05}{51}{46}{254}$,
$\delta_{(2000)}=\dec{+02}{48}{29}{65}$ \citep{Rodriguez2008}
and
$\alpha_{(2000)}=\ra{05}{51}{46}{07}$,
$\delta_{(2000)}=\dec{+02}{48}{30}{76}$, respectively
\citep{Reipurth1999,Rodriguez2008}.
The jet is bipolar with a western and an eastern component. It
has a position angle (P.A.) of $\sim$ 97\degree{}
and an inclination angle of $\sim$ 10\degree{}, with the western component
being blueshifted and tilted toward us \citep{Reipurth1992}.
The systemic velocity in this system is assumed to be 8.90$\pm$0.14 \vkm{}
LSR, as before. Throughout this paper, the velocity is relative to this systemic value.

\subsection{1.33 mm Continuum Emission} \label{ssec:cont}

%The emission is compact and no emission is detected further out from the
%VLA 1 source along the equatorial plane likely due to not enough sensitivity.
% and the gas and dust associated with
%this emission have a mass of $\sim 0.13 \pm 0.03$ \solarmass{}. 

As shown in Figure \ref{fig:cont}a, the continuum emission associated with
the VLA 1 source is now resolved into a disk-like structure perpendicular to
the jet axis, with a Gaussian deconvolved (FWHM) size of $\sim$
\arcsa{0}{6}$\times$\arcsa{0}{3} ($\sim$ 240$\times$120 AU)
and a P.A. of 7.6\degree{}. 
As found in \citet{Lee2010HH111}, this
emission has a total flux of
$\sim$ 285$\pm$40 mJy and it is the thermal emission mainly from a dusty
disk, with the inner part of which already seen with a deconvolved size of
$\sim$ \arcsa{0}{15} (60 AU) at 7 mm \citep{Rodriguez2008}. 
Note that the emission seems to
show a little faint protrusion to the east along the jet axis.
%Note that the emission is not resolved in the minor axis and it seems to
%show a little faint protrusion to the east along the jet axis. 
This faint
protrusion seems to be seen in 3.6 cm and 7 mm as well, and it may trace
either the material along the jet axis or a disk around a companion in a
close binary system as suggested in \citet{Rodriguez2008}. In addition, a
faint protrusion is also seen extending to the southeast from the southern
part of the disk.
% and it could trace the envelope around the cavity wall.
As in \citet{Lee2010HH111}, faint emission is also detected around the
VLA 2 source ($\sim$ 4 $\sigma$ detection with 1$\sigma=0.9$
\mJyb{}), probably tracing a disk around it. Moreover, two more faint emission peaks
DCN and DCS are seen at \arcs{3} in the north and \arcs{2} in the south
respectively near the equatorial plane (P.A. $\sim$ 7\degree{}), probably
tracing the density enhancement there.

%On the other hand, the total flux toward the VLA 2
%source is $\sim$ 11$\pm2$ mJy. The mass would be $\sim$ 0.005 \solarmass{}
%if the emission has the same properties as that around the VLA 1 source.

\subsection{Previous Results of Envelope and Outflow in \cCO{}}

%, with the innermost part overlapping
%with the continuum emission associated with the VLA 1 source. 

%The envelope seems to also have some infall motion that is smaller than the
%rotation motion. Thus, the material in the outer part of the envelope seems
%to be slowly spiraling inward with its angular momentum. and the rotation
%may become Keplerian in the inner part.

Previously in the \cCO{} observations \citep{Lee2010HH111}, a flattened
envelope is seen perpendicular to the jet axis, extending to $\sim$ 7200 AU
(\arcs{18}) out from the VLA 1 source. It is rotating, with the blueshifted
emission in the north and the redshifted emission in the south. The outer
part ($\sim$2000$-$7200 AU, or $\sim$
\arcs{5}--\arcs{18}) seems to have a rotation that has constant
specific angular momentum with $v_\phi\sim 3.90 d^{-1}$ \vkm{}, 
while the inner part ($\sim$ 60$-$2000 AU, or $\sim$
\arcsa{0}{15}--\arcs{5}) seems to have a Keplerian rotation with $v_\phi
\sim 1.75 d^{-0.5}$ \vkm{}, where $d$ is the radial distance from the source
in the unit of arcsecond. Outflow is also seen in \cCO{}, with the
blueshifted emission extending to the west and the redshifted emission
extending to the east.

%In addition, the blueshifted emission shows a
%V-shaped structure opening to the north, spatially coincident with the
%cavity walls traced by the reflection nebulae.
%suggesting that the envelope
%material is piling up on the cavity walls as the cavity walls expand into
%the envelope.

\subsection{Line Emission} \label{ssec:line}

The \bCO{} and \aCO{} lines are detected toward the VLA 1 source upto $\sim$
4.2 \vkm{} and 10 \vkm{}, respectively, from the systemic velocity. In order
to show how the envelope and disk structure changes with velocity, the line
emission is divided into four velocity ranges: low (0$-$2.5 \vkm{}), medium
(2.5$-$4.2 \vkm{}), high (4.2$-$5.8 \vkm), and very high (5.8$-$10 \vkm{})
velocity ranges, on the redshifted and blueshifted sides.

\begin{figure} [!hbp]
\centering
\includegraphics[angle=-90,scale=0.7]{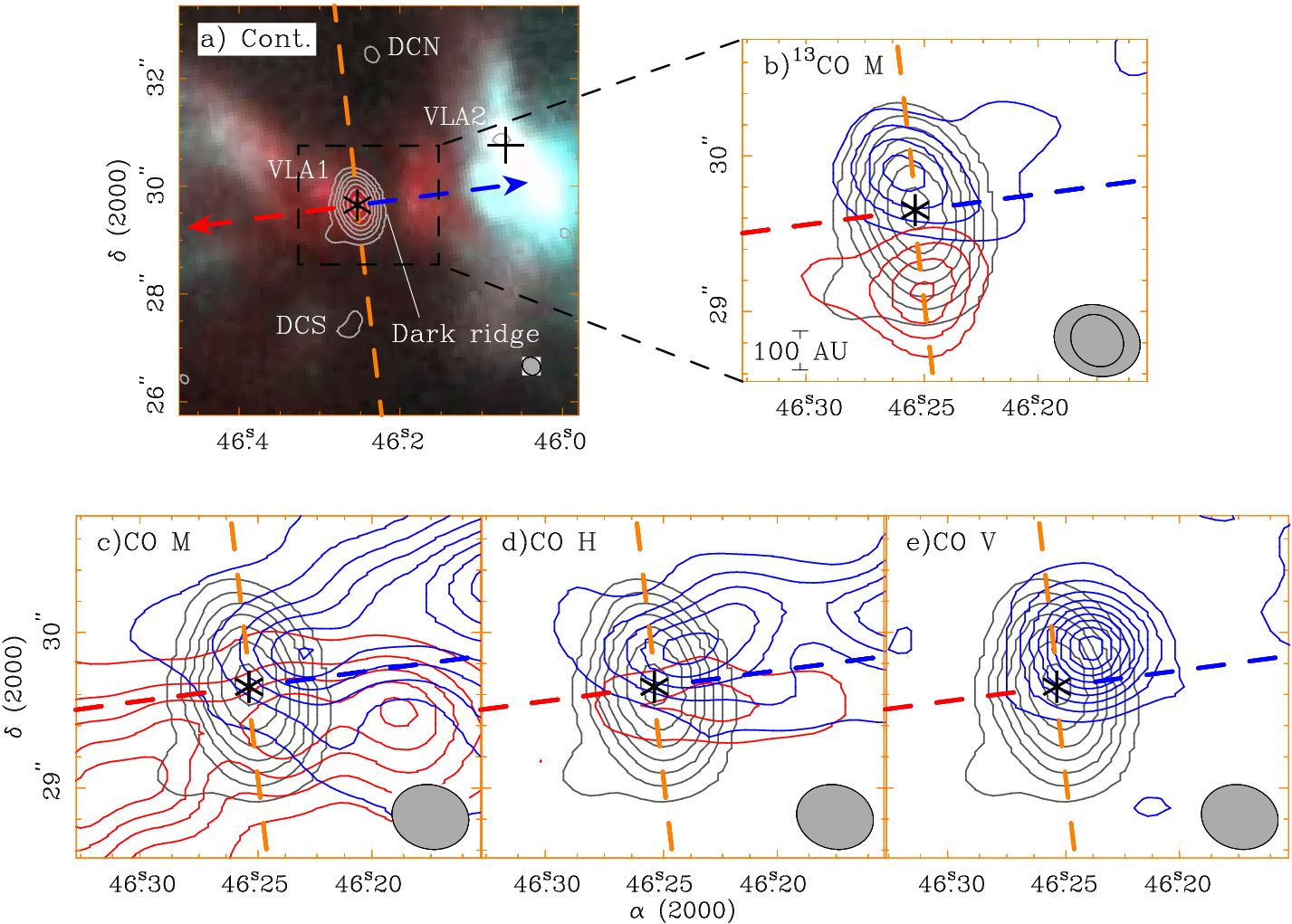}%{cont.ps}
\figcaption[]
{1.33 mm continuum, \bCO{}, and \aCO{} emission contours overplotted on the
{\it HST} NICMOS image adopted from \citet{Reipurth1999}.  The asterisk and cross mark
the positions of the VLA 1 and 2 sources, respectively. 
The blue and red arrows indicate the
orientations of the blueshifted and redshifted components of the jet,
respectively.
The dashed line indicates the equatorial plane perpendicular to the jet axis.
The synthesized beam has a size of \arcsa{0}{36}$\times$\arcsa{0}{31} in
continuum, \arcsa{0}{66}$\times$\arcsa{0}{56} in \bCO{}, and 
\arcsa{0}{58}$\times$\arcsa{0}{50} in \aCO{}.
The gray contours are
the continuum contours with levels of $4\sigma (1-r^n)/(1-r)$, where $r=1.45$,
n=1,2,3.., and $\sigma=0.9$ \mJyb{}.
%from 4 to 124$\sigma$ with a step of 10$\sigma$,
%where $\sigma= 0.9$ \mJyb{}.
(b) shows the redshifted (red contours) and blueshifted (blue contours)
\bCO{} emission at medium velocity. The contour levels start at 4$\sigma$
with a step of 3$\sigma$, where $\sigma=0.023$ \Jyb{}.
(c), (d), and (e) show the
redshifted (red contours) and blueshifted (blue contours)
\aCO{} emission at medium velocity, high velocity,
and very high velocity, respectively.
The contour levels start at 4$\sigma$
with a step of 6$\sigma$ in (c), and a step of 4$\sigma$ in (d) and (e),
where $\sigma=0.025$ \Jyb{}.
\label{fig:cont}
}
\end{figure}

\begin{figure} [!hbp]
\centering
\includegraphics[angle=-90,scale=1]{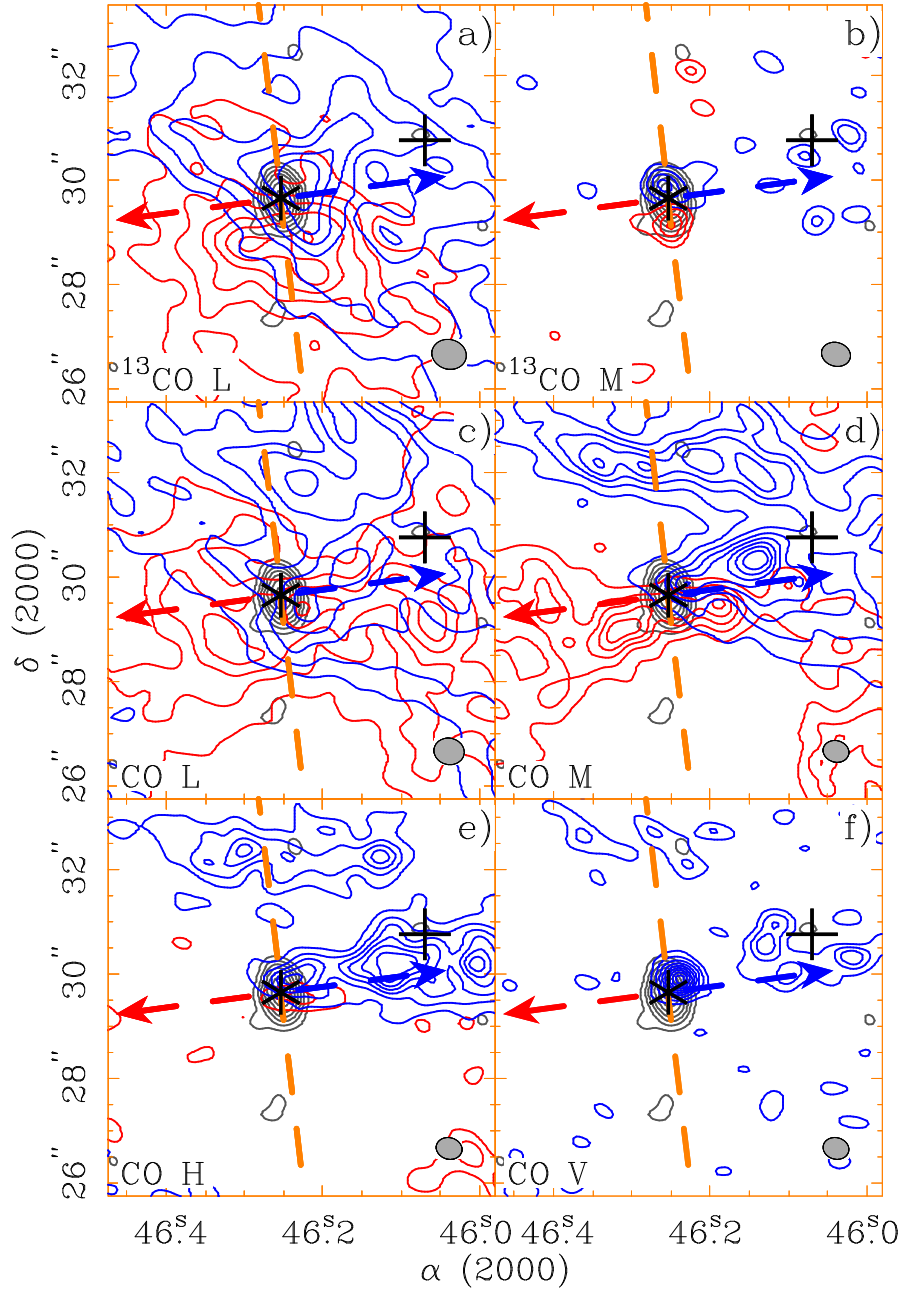}%{cont_l.ps}
\figcaption[]
{Same as Figure \ref{fig:cont} but
adding low-velocity \bCO{} and \aCO{} emission contours,
without zooming into the central region.
Here (b) corresponds to Figure \ref{fig:cont}b, and (d)-(f) correspond to
Figure \ref{fig:cont}c-e. (a) shows the
the redshifted (red contours) and blueshifted (blue contours) \bCO{} emission
at low velocity. The contour levels start at 4$\sigma$ with a step of
6$\sigma$, where $\sigma=0.023$ \Jyb{}.
(c) shows the
the redshifted (red contours) and blueshifted (blue contours) \aCO{} emission
at low velocity. The contour levels start at 4$\sigma$ with a step of
8$\sigma$, where $\sigma=0.025$ \Jyb{}.
\label{fig:line}
}
\end{figure}

In \bCO{}, at low velocity, the redshifted emission is detected mainly in
the south and the blueshifted emission is mainly in the north of the VLA 1
source (Fig. \ref{fig:line}a), consistent with a rotation motion about the
source. As seen in \cCO{}, the emission is contaminated by the outflow emission, with the
blueshifted emission also extending to the west and the redshifted emission
also extending to the east. In addition, the blueshifted emission shows a
V-shaped structure opening to the north, spatially coincident with the
cavity walls traced by the reflection nebulae, suggesting that the envelope
material is piling up on the cavity walls as the cavity walls expand into
the envelope. A similar V-shaped structure is also seen opening to the south
in the redshifted emission. At medium velocity, the \bCO{} emission shrinks
to the source, spatially coincident with the dusty disk (Fig.
\ref{fig:line}a and Fig. \ref{fig:cont}b for a zoom-in). The blueshifted
emission peak is to the north and the redshifted emission peak is to the
south in the equatorial plane, indicating that the motion there is highly
dominated by the rotation about the VLA 1 source, as expected for a
rotationally supported disk. The two peaks, however, are not symmetric about
the source, with the northern peak at a distance of
$\sim$ \arcsa{0}{25} and southern peak at $\sim$ \arcsa{0}{5}. Two faint
protrusions are also seen, one in the blue extending to the west from the
northern part of the disk and one in the red extending to the east from the
southern part, parallel to the jet axis. These structures are similar to
those seen in the rotating molecular outflow in CB 26 \citep{Launhardt2009},
and thus may trace material outflowing from the disk.

%the emission of the envelope is mostly resolved
%out by the SMA (Fig. \ref{fig:line}c). Nonetheless, it is still clear that

In \aCO{}, at low velocity, the blueshifted emission is mainly to the north
and the redshifted emission is mainly to the south, with V-shaped structures
coincident with the cavity walls (Fig. \ref{fig:line}c), similar to that
seen in \bCO{}. At medium velocity, the blueshifted emission extends to the
west from the northern part of the disk while the redshifted emission
extends to both the west and east from the southern part of the disk (Fig.
\ref{fig:line}d and Fig. \ref{fig:cont}c for a zoom-in). Note that the
blueshifted emission and the redshifted emission
beyond $\sim$ $\pm$\arcs{1} from the source are
contaminated by the outflow shell emission, showing a cone-like structure
with the tip pointing toward the source. The blueshifted emission also shows
an E-W elongation at $\sim$
\arcs{3} in the north (Fig. \ref{fig:line}d), with an unknown origin.
At high velocity, the blueshifted and redshifted emissions extend out
from the inner radii
of the disk (Fig. \ref{fig:line}e and Fig. \ref{fig:cont}d for a zoom-in).
The emissions within $\sim$ \arcs{1} from the source are not much contaminated by
the outflow shell emission, which is expected to be further away from the
source at higher velocity. They extend to the west, first opening from the
VLA 1 source and then bent to be aligned with the jet axis. They also extend
slightly to the east. The blueshifted emission is mainly to the
north of the jet axis and the redshifted emission is mainly to the south,
with the velocity sense the same as that of the rotation in the disk, and
thus may trace the material coming out from the disk. Note that,
in the west, the blueshifted emission also extends slightly to the south. At
very high velocity, only blueshifted emission is detected with a peak
slightly away from the equatorial plane at $\sim$ \arcsa{0}{25} to the west (Fig.
\ref{fig:line}f and Fig. \ref{fig:cont}e for a zoom-in). This emission peak
is spatially coincident with the blueshifted emission peak seen at high
velocity in Figure \ref{fig:cont}d, suggesting that the emission there has a
broad range of velocities from high to very high and may trace a shock
interaction, either inside the disk or outside.

%Higher resolution observation
%is needed to separate this emission from that of the disk.
%This transverse elongation seems not trace the cavity wall which is a bigger
%opening angle.
%Could this elongation perpendicular arise from the envelope that produces
%the HH 121 jet, since the direction is almost perpendicular to the jet?

\subsubsection{Rotation motion near the VLA 1 source}\label{ssec:rotation}

%As mentioned above, the blueshifted emission is more affected by outflow
%interaction, while the redshifted emission is less. Thus, we will focus more
%on the redshifted side when deriving the rotation velocity.

%Previously in the \cCO{} observations, the envelope is found to have a
%differential rotation, with the outer part ($\sim$ \arcs{5}--\arcs{18})
%better described by a rotation that has constant specific angular momentum
%and the inner part ($\lesssim$ \arcs{5}) by a Keplerian rotation
%\citep{Lee2010HH111}. 

\begin{figure} [!hbp]
\centering
\includegraphics[angle=270,scale=0.80]{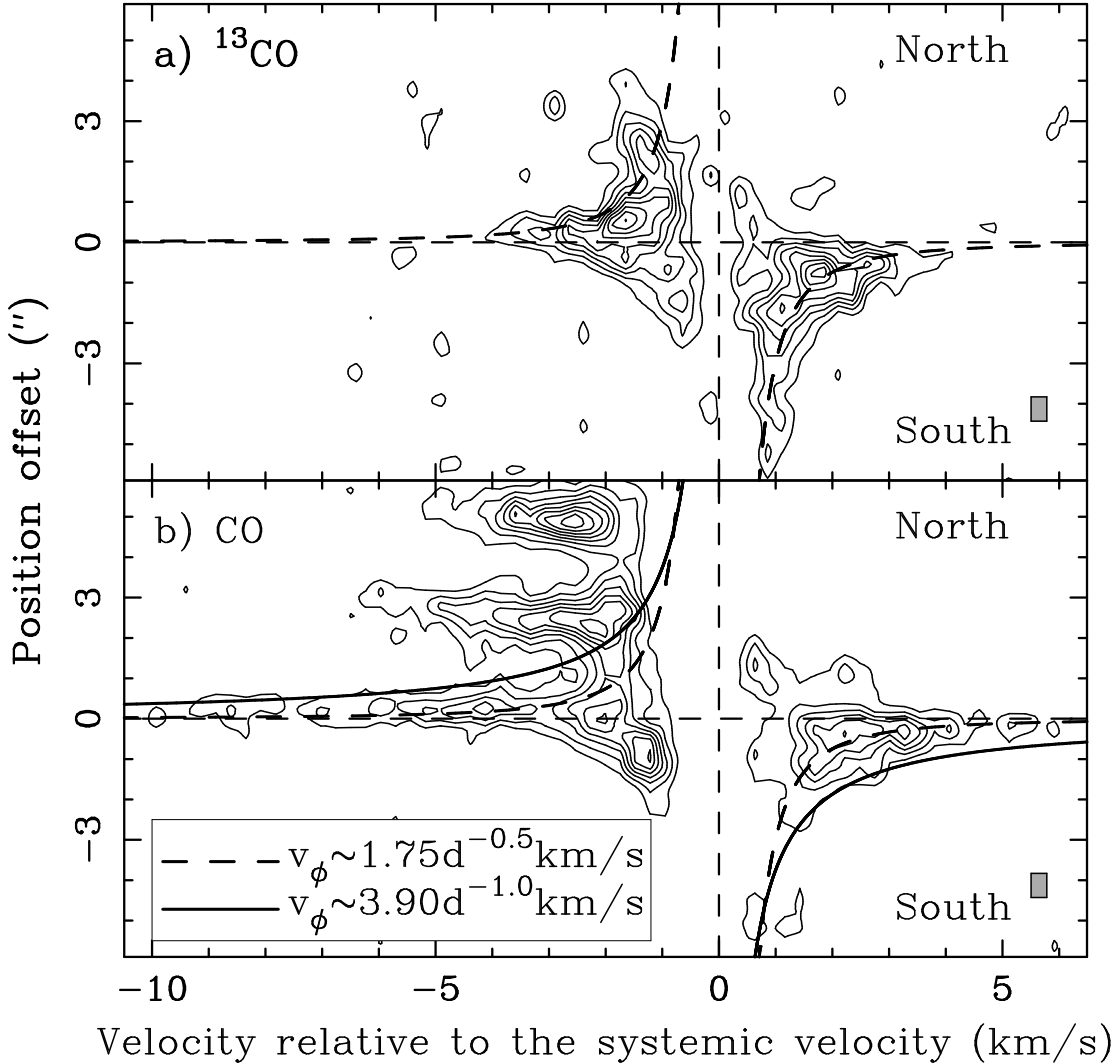}%{pv.ps}
\figcaption[]
{Position-velocity (PV) diagrams in (a) \bCO{}
and (b) \aCO{} centered at the VLA 1 source cut along the equatorial plane
perpendicular to the jet axis.
The gray box in the lower-right corner shows the velocity and
spatial resolutions of the PV diagrams.
Solid curve shows the rotation that has constant specific
angular momentum and dashed curve shows the Keplerian rotation, as
derived from the \cCO{} envelope in \citet{Lee2010HH111}, with
both corrected for the inclination angle.
Here $d$ is the radial distance from the source in arcsec.
In (a), the contour levels start at 3$\sigma$ with a step of
3$\sigma$, where $\sigma=0.023$ \Jyb{}.
In (b), the contour levels start at 3$\sigma$ with a step of
4$\sigma$, where $\sigma=0.025$ \Jyb{}.
\label{fig:pvrot}
}
\end{figure}

As seen in the last section, the inner part of the envelope can also be
traced by \bCO{} and \aCO{}, allowing us to check and confirm the Keplerian
rotation law near the source seen in \cCO{}. Figure
\ref{fig:pvrot} shows the position-velocity (PV) diagrams of the envelope
and disk in \bCO{} and \aCO{} cut along the equatorial plane perpendicular
to the jet axis. In \bCO{}, the emission within $\pm$1 \vkm{} from the
systemic velocity is mainly resolved out and affected by the outflow
emission, and thus should be excluded for studying the rotation law. As can
been seen, the velocity structure beyond $\pm$1
\vkm{} roughly follows the Keplerian rotation law seen in \cCO{}.
In \aCO{}, the emission within $\pm$2 \vkm{} from the systemic velocity is
mainly resolved out and affected by the outflow emission, and thus should
also be excluded for studying the rotation law. 
In addition, as seen in the emission map
(Fig. \ref{fig:line}d and e), the blueshifted emission is also detected at
$\sim$ \arcs{2} to \arcs{3} in the
north with a velocity ranging from $-$2 to $-$6 \vkm{} w.r.t. the systemic
velocity. This emission is unclear, and it may trace some
interaction in the envelope or something in the foreground or background,
and thus should also be excluded.
In this line, on the
redshifted side, the emission is detected up to $\sim$ 6 \vkm{}, higher than
that seen in \bCO{}, arising from the inner region as seen in Figure
\ref{fig:cont}d. The velocity structure also follows the Keplerian rotation
law to that high velocity to the inner region. On the blueshifted side, the
emission is detected up to $-$10 \vkm{} from the systemic, much higher than
that on the redshifted side. However, as can be seen, the emission beyond
$-$5 \vkm{} is localized with a broad range of velocities and it corresponds
to the localized emission at $\sim$
\arcsa{0}{25} to the west as seen in the emission map (Fig. \ref{fig:cont}e). Since
this emission can trace a local shock interaction, the blueshifted emission
should not be used to derive the rotation law. As
a result, the rotation structure near the source seen in \aCO{}
is also consistent with that seen in \cCO{}.

% {\bf which 
%results in a mass of 1.25 \solarmass{} for the central source, after
%subtracting the mass of the dusty disk close to the source.}

%As discussed above and also can be seen in the following, the \bCO{} and \aCO{}
%emission around the systemic velocity 
%is not only mostly resolved out and self absorbed because of the high abundance of
%these molecules, but also affected by the outflow emission.
%%that trace the more extended structure
%Thus, only the emission at the high velocity end can be used to study the
%rotation motion. The emission there traces the envelope and disk near the
%source.

\section{Model the disk continuum emission} \label{sec:model}

Here a simple disk model is used to fit the structure and intensity
distribution of the continuum emission in order to derive the temperature
and density distributions of the dusty disk. Note that, the emission that
extends to the east along the jet axis and the emission that extends to the
southeast from the southern part of the disk are not from the disk
perpendicular to the jet axis and are thus excluded from the model.

%This power-law index is also expected for an optically thin disk
%\cite{Garaud2007}, as the case for HH 111 disk \citep{Lee2009HH111}.

%With $q=0.5$, the temperature reaches
%64 K at \arcsa{0}{15} (or 60 AU), unresolved in our observation.
%78 K at \arcsa{0}{1}

%The observed continuum emission structure along the
%major axis of the disk is slightly asymmetric about the source position, and
%the model here is geared to match the structure and intensity distribution
%in the south.

\begin{figure} [!hbp]
\centering
\includegraphics[angle=-90,scale=0.67]{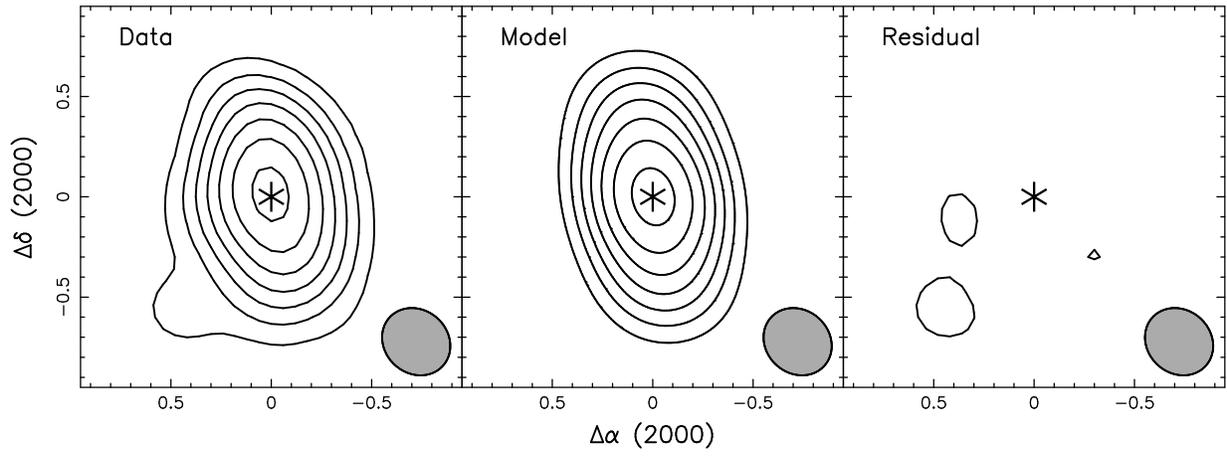}%{fitcont.ps}
\figcaption[]
{A fit to the continuum emission with a simple disk model as described in
Section \ref{sec:model}. 
%The gray and black contours show the observed and
%model continuum emission, respectively. 
The asterisk marks the position of
the VLA 1 source. The contour levels are the same as those in Figure
\ref{fig:cont}a.
\label{fig:fitcont}
}
\end{figure}

For simplicity, the disk in the model is assumed to be flat with a constant
thickness of $H$, an inner radius of $\Rin$, and an outer radius of $\Rout$.
The disk is assumed to be perpendicular to the jet axis and thus have an
inclination angle of 10\degree{} to the line of sight, with the nearside
tilted to the east \citep{Reipurth1992}. It has a position angle of
7.6\degree{} as given by the Gaussian source deconvolution earlier in Section
\ref{ssec:cont}. The
temperature is assumed to be given by
\begin{equation}
T=T_0 (\frac{R}{R_0})^{-q}
\end{equation}
where $R_0=\arcsaq{0}{3}$ (120 AU), which is the radius (i.e., the half of
the size) of the disk derived earlier from the Gaussian source
deconvolution, $T_0$ is the temperature at
$R_0$, and the power-law index $q$ is set to 0.5, assumed to be similar to that found in T-Tauri disks in
the late stage of star formation
\citep{Mundy1996,Andrews2009}. In \citet{Lee2009HH111}, 
using the dust opacity law 
\begin{equation}
\kappa_\nu = 
0.1 \left( \frac{\nu}{10^{12}\textrm{Hz}} \right) ^\beta 
\;\;\textrm{cm}^2 \;\textrm{g}^{-1}
\end{equation}
 as given in \citet{Beckwith1990},
the spectral energy
distribution (SED) of the continuum source was found to be fitted with a
dust temperature ranging from 41 to 64 K and $\beta\sim 0.9$.
Thus, here $T_0$ is assumed to be 45 K and
the dust mass opacity is assumed to be 0.026 cm$^2$ g$^{-1}$ at
$\lambda=1.33$ mm (or 225 GHz).
The number density of molecular hydrogen is assumed to be given by
\begin{equation}
n=n_0 (\frac{R}{R_0})^{-p}
\end{equation}
with the power-law index $p$ assumed to be 1.
With a constant thickness, the
disk has a surface density with a power-law index of 1, similar to that
found in the T-Tauri disks \citep{Andrews2009}. 
As can be seen in Figure \ref{fig:fitcont}, with the above assumptions, the model can fit the continuum emission
reasonably well, with $H \sim \arcsaq{0}{3}$ (120 AU),
$\Rin \sim \arcsaq{0}{03}$ (12 AU), $\Rout \sim \arcsaq{0}{6}$ (240 AU), and
$n_0 \sim 10^9$ \cmc{}.
Here, $H$ turns out to be the same as the Gaussian source
deconvolved size of the continuum emission in the minor axis.
This is expected for a nearly edge-on disk, for which the size in the minor
axis is mainly due to the thickness of the disk.
%Note that, however, the disk is unresolved in the minor axis and could be thinner.
On the other hand, $\Rout \sim 2 R_0$ and thus is twice the disk radius derived
from the Gaussian source deconvolution. 
Assuming that Helium has a number density
$n_\textrm{\scriptsize He}=0.2n$, then the mean molecular weight with
respect to the number of molecular hydrogen is 2.8.
Thus, the mass of the disk is found to be
0.14 \solarmass{}, similar to the mean value found by fitting the SED
\citep{Lee2010HH111}. The surface density in the disk is
$\sim$ 7.2 g cm$^{-2}$ at 120 AU and 21.5 g cm$^{-2}$ at 40 AU, similar to
those found in the bright T-Tauri disks, e.g., GSS 39, in Ophiuchus, which
also has a similar radius of $\sim$ 200 AU and a same disk mass of $\sim$
0.14 \solarmass{} \citep{Andrews2009}. Thus, the dusty disk here can be a
very young protoplanetary disk that could evolve into a T-Tauri disk in the
later stage of star formation.

%This difference is a consequence of
%the central concentration in the emission caused by the temperature and
%surface density gradients in the power-law disk model as discussed in
%\citet{Mundy1996}. 

%In Andrews et al. Rc seems to be more similar to the outer radius.
%The disk surface density varying roughly inversely with radius in the
%inner disk (\propto 1/R when R \lesssim Rc) and smoothly merging into a
%steeper exponential decrease at larger radii (1/e^R when R \gtrsim Rc). The
%characteristic radius Rc mark that transition.

%The temperature here is much higher than that can be provided by the
%viscous accretion disk, which predict about 11 K at 0.3", assuming an
%accretion rate of 4.2e-6 and a stellar mass of 1.25. Thus, irradiation by
%the central star is important, as in the T-Tauri disks.

\section{Discussion}

\subsection{A Rotationally supported Disk}

% Note that the disk could be larger in these gas tracers.

A rotationally supported disk has been seen in Taurus with an outer radius
of $\sim$ 500 AU in CO in the late (i.e., Class II or T Tauri) phase of star
formation \cite[see, e.g.,][]{Simon2000}. Note that the disk could appear
smaller in continuum \citep{Andrews2009}, because of the rapid decrease in
the continuum intensity due to a power-law decrease of the density and
temperature in the disk. Formation of such a disk must have started early in
the Class I phase and even earlier in the Class 0 phase. HH 111, with a
bolometric temperature of only $\sim$ 78 K, is a very young Class I source, just
transitioning from the Class 0 phase.
%As argued in
%\citet{Lee2010HH111}, a rotationally-supported disk must have already
%formed around the source to launch the jet in the HH 111 system. 
The observations here show that such a rotationally supported disk can
indeed form in such an early phase of star formation. In continuum, the disk
is found to have a deconvolved radius of $\sim$ 120 AU (with an outer radius
of 240 AU in the disk model), with the inner part of which already seen with
a deconvolved radius of $\sim$ 30 AU at 7 mm \citep{Rodriguez2008}. 
It is
rotationally supported with a Keplerian rotation, as seen previously in
\cCO{} and now in \bCO{} and \aCO{}, which implies a mass of 
$\sim$ 1.3 \solarmass{} for the central star \citep{Lee2010HH111}.
Note that as discussed in \citet{Lee2010HH111}, this stellar mass
can account reasonably for the
observed bolometric luminosity.
The disk mass is $\sim$ 0.14 \solarmass{}, which is $\sim$ 10\% of
the stellar mass.
This disk mass is consistent with that implied for a Class 0 source by
\citet{Jorgensen2009} and that implied for a young Class I source by
\citet{Andrews2007}.
Since no rotationally supported disk  has been confirmed in the Class 0 phase,
the disk here could be the youngest rotating disk resolved to date.
Moreover, as discussed earlier, the surface density and the radius of the
disk here in HH 111 are both similar to those of the bright resolved T-Tauri
dusty disks in \citet{Andrews2009}. Thus, the disk could evolve to a
T-Tauri disk in the later stage of star formation as the accretion rate
drops to 10$^{-7}$ \smassrate{} from a few times 10$^{-6}$ \smassrate{}
\citep{Lee2010HH111}.

As discussed in \citet{Lee2010HH111}, the Keplerian rotation in this system
can extend
further out to the envelope to $\sim$ 2000 AU (\arcs{5}) from the source. At that
large radius, the envelope seems to have some small amount of infall
motion as well, and thus the rotation there could actually be sub-Keplerian.
Therefore, the disk seems to be deeply embedded in a sub-Keplerian envelope
with a radius of $\sim$ 2000 AU. 
Observations at higher velocity and angular
resolutions are really needed to differentiate the Keplerian motion from the
sub-Keplerian motion in order to determine where the envelope is
transitioning to a disk. Another way to determine the location of transition
is to search for an accretion shock that is expected to occur at the
transition. As discussed in
\citet{Lee2010HH111}, SO is found to trace a shock emission at 400 AU
(\arcs{1}), and it may trace an accretion shock and thus the transition
there. Further observations in other shock tracers are needed to check this
possibility.

Rotationally supported disks have also been claimed in a few other Class I
sources, e.g., IRAS 04302+2247 (the Butterfly star, $\Tbol=203$ K)
\citep{Wolf2008}, L1489-IRS ($\Tbol=238$ K) \citep{Brinch2007}, IRS 43
($\Tbol=310$ K)
\citep{Jorgensen2009}, and Elias 29 ($\Tbol=350$ K) \citep{Lommen2008}.
These sources, with a much higher $\Tbol$ than HH 111, are at much later
stages than HH 111. Among these sources, L1489-IRS has been modeled in
detail for the disk and envelope properties, and thus can be compared here
with HH 111. This system was found to host a Keplerian disk with an outer
radius of
$\sim$ 200 AU deeply embedded in a sub-Keplerian envelope with a radius
$\sim$ 2000 AU \citep{Brinch2007}, similar to the HH 111 system. The mass 
and the infall rate in
the sub-Keplerian envelope are
$\sim$ 0.093 \solarmass{} and $\sim$ 4.3$\times10^{-6}$ \smassrate{},
respectively \citep{Brinch2007}, also similar to those in HH 111, which are
$\sim$ 0.13 \solarmass{} and 4.2$\times10^{-6}$ \smassrate, respectively
\citep{Lee2010HH111}.
The central star has a mass of $\sim$ 1.35 \solarmass{}, also similar to that in
HH 111. However, this system has a bolometric luminosity of $\sim$ 3.7
\solarlum{}, much smaller than that of HH 111, which is $\sim$ 20
\solarlum{}.  The small bolometric luminosity suggests that the accretion
rate in the disk must be much lower than that in HH 111, as supported by the
less jet activity. This is probably because that the accretion rate decreases
with the evolutionary stage. The disk mass was first found to be only
4$\times 10^{-3}$
\solarmass{} \citep{Brinch2007} and later modified to be 
0.018 \solarmass{} \citep{Jorgensen2009}, but it is still much smaller than
that in HH 111. As argued in \citet{Jorgensen2009}, however, it is not clear
if this is because the disk mass really decreases with the evolutionary
stage, or merely because the disk mass is underestimated more in the later
stage due to a grain growth.

%Similar trend is also found by \citet{Jorgensen2009} and
%\citet{Andrews2007}.

Rotationally supported disks have been seen in a
few other much older sources, eg., HH 30 (late Class I)
\citep{Guilloteau2008} and IRAS 04158+2805 (Class II)
\citep{Andrews2008}. In these older disks, a central hole is seen and it
could be due to a tidal clearing by a possible binary system at the center or
due to an outflow cavity.
Since the HH 111 VLA 1 source itself is a close binary, a central hole is
expected to have formed at the center of the disk near the
source as well. Observations at higher angular resolution are needed to
check this possibility.

%Bright dusty disks can be reasonably resolved in the later stage
%of star formation in the Class II phase in Ophicious with a characteristic
%radius of upto $\sim$ 200 AU \citep{Andrews2009}.
%Keplerian rotation are already detected in CO J=2-1 and the gaseous
%usually can extend further out to much larger radius, with a outer radius
%up to 800 AU \citep{Simon2000}.

\subsection{Material outflowing from the disk?}\label{ssec:rotation_out}

%At a cut 1.2" away from the
%source, the blueshifted emission extends to the south of the jet axis and
%extends to higher blueshifted velocity than that in a), probably because of
%an additional blueshifted velocity due to the outflow motion parallel to the
%jet axis, or the outflow contanimation that shows a fan-shaped structure
%extending to the west. For the redshifted side, however, it is not clear if
%the emission extends more to the north when we go from cut D to cut A. Thus,
%it is unclear if there is systemic velocity shift between the cut in the
%east and the cut in the west. 

\begin{figure} [!hbp]
\centering
\includegraphics[angle=-90,scale=0.85]{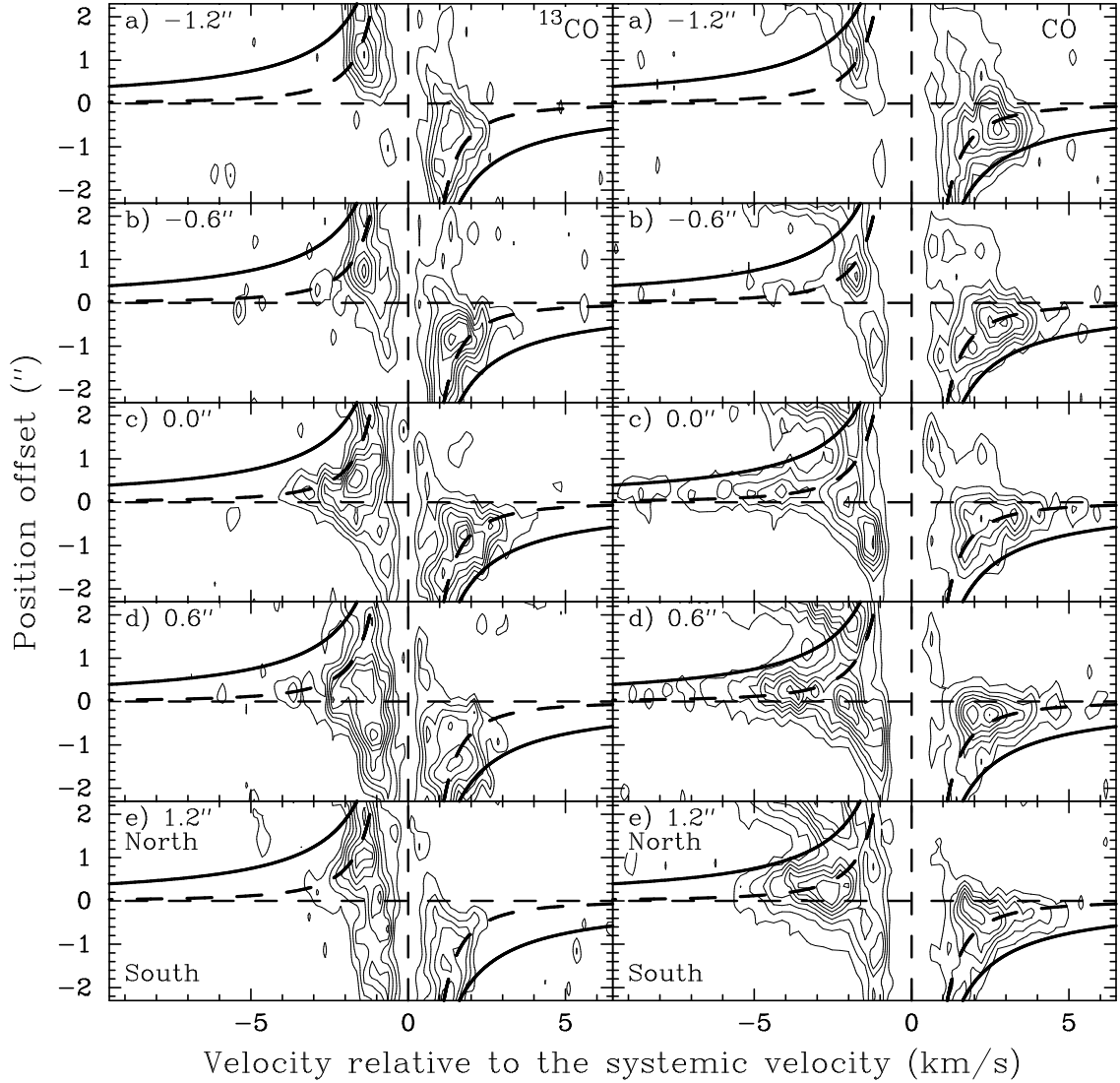}%{pvj.ps}
\figcaption[]
{PV diagrams in \bCO{} (left column) and \aCO{} (right column) cut across
the jet axis centered at 5 positions along the jet axis, from (a)
$-$\arcsa{1}{2} in the east to (e) \arcsa{1}{2} in the west with a step of
\arcsa{0}{6} (which is the size of the synthesized beam).  The gray box in
the lower-right corner of (a) shows the velocity and spatial resolutions of
the PV diagrams. The solid and dashed curves, and contour levels are the
same as those in Figure
\ref{fig:pvrot}.
\label{fig:pvj}
}
\end{figure}

As mentioned in Section \ref{ssec:line}, \aCO{} and \bCO{} emissions are
seen near the source extending out from the disk to the east and west
parallel to the jet axis. Could this emission structure trace the material
coming out from the disk, as that seen in CB 26 \citep{Launhardt2009},
carrying away the angular momentum? To study this possibility, PV diagrams
cut across the jet axis centered at 5 positions along the jet axis, from
$-$\arcsa{1}{2} in the east to \arcsa{1}{2} in the west with a step of
\arcsa{0}{6} (which is the size of the synthesized beam), are presented here
in Figure \ref{fig:pvj} to investigate the kinematics of this emission
structure. In these cuts, excluding the emission within $\pm$2 \vkm{} from
the systemic velocity, the blueshifted emission is in the north and
redshifted in the south, as seen in the emission map in Figure
\ref{fig:cont}d, with a velocity sense the same as that of the rotation in
the disk. Therefore, the emission could trace the material rotating around the
jet axis and thus the material outflowing from the disk. In panels (c) and (d), the blueshifted
emission is contaminated by the shock emission and should be excluded. From
cut A in the east to cut E in the west, no clear systemic velocity shift is
seen in the PV diagrams. Thus, if the emission really traces the material
outflowing from the disk, the projected outflow velocity must be smaller
than 1 \vkm{}.  This is not impossible since (1) the outflowing material
could have the same inclination as the jet and thus be close to the plane of
the sky with an inclination angle of $\sim$ 10\degree{}, and (2) the
outflowing velocity could be only a few \vkm{} in \aCO{}
\citep{Launhardt2009}. Further observations are needed to study this
possibility.

%Alternatively, the
%emission could just trace the envelope material.

\section{Conclusion}

The HH 111 protostellar system has been mapped at a resolution of $\sim$
\arcsa{0}{3} in 1.33 mm continuum, and \arcsa{0}{6} in \bCO{} ($J=2-1$) and
\aCO{} ($J=2-1$). The 1.33 mm continuum emission shows a resolved dusty disk
associated with the VLA 1 source perpendicular to the jet axis, with a
Gaussian deconvolved size of $\sim$ 240 AU. The \bCO{} and \aCO{} emissions
toward the dusty disk show a Keplerian rotation, indicating that the dusty
disk is rotationally supported. The density and temperature distributions in
the disk derived from a simple disk model are found to be similar to those
found in bright T-Tauri disks. Thus, the disk can be a young protoplanetary
disk that could evolve into a T-Tauri disk in the late stage of star
formation. In addition, a hint of a low-velocity molecular outflow is also
seen in \bCO{} and \aCO{} coming out from the disk.

\acknowledgements

I thank the SMA staff for their efforts in running and maintaining the
array. I also thank the referee for the valuable comments.

\clearpage

\end{document}